\documentclass[12pt,aps,prd,a4paper,titlepage,preprint,tightenlines,superscriptaddress,showpacs,nofootinbib,floatfix]{revtex4}

\pdfoutput=1
\usepackage{url,amsmath,amssymb,slashed}

\usepackage[colorlinks=true,citecolor=blue,urlcolor=blue]{hyperref}

  \usepackage{graphicx}                    

\newcommand{\vc}[1]{\boldsymbol{#1}}
\newcommand{\s}[1]{_{\rm #1}} 
\newcommand{\software}[1]{\texttt{#1}}
\makeatletter
\protected\def\EE{\@ifnextchar-{\@@EE}{\@EE}}
\protected\def\@EE#1{\ifnum#1=1 \!\times\!10 \else \!\times\!10^{#1}\fi}
\protected\def\@@EE#1#2{\!\times\!10^{-#2}}
\makeatother
\newcommand{\un}[1]{\,{\rm #1}}
\newcommand{\TeV}{\un{TeV}}
\newcommand{\GeV}{\un{GeV}}
\newcommand{\MeV}{\un{MeV}}

\newcommand{\ab} {\un{ab}}
\newcommand{\iab}{\un{ab^{-1}}}

\renewcommand{\eqref}[1]{Eq.~(\ref{#1})}

\newcommand{\secref}[1]{Sec.~\ref{sec:#1}}
\newcommand{\secsref}[2]{Secs.~\ref{sec:#1} and \ref{sec:#2}}
\newcommand{\figref}[1]{Fig.~\ref{fig:#1}}
\newcommand{\Figref}[1]{Figure~\ref{fig:#1}}
\newcommand{\figsref}[2]{Figs.~\ref{fig:#1} and \ref{fig:#2}}
\newcommand{\tableref}[1]{Table~\ref{table:#1}}

\let\gsim\gtrsim
\let\lsim\lesssim

\newcommand{\mET}{{\slashed E}\s{T}}
\newcommand{\PT}{P\s{T}}
\newcommand{\Eloss}{E\s{loss}}
\newcommand{\gravitino}{\tilde{G}}

\newcommand{\mplanck}{M\s{Pl}}

\newcommand{\dd}{{\rm d}}

\newcommand{\Normal}{\mathop{\mathrm N\phantom{}}}
\newcommand{\beq}{\begin{equation}}
\newcommand{\eeq}{\end{equation}}
\begin{document}
\makeatletter%
\def\PRE#1{\preprintsty@sw{#1}{\relax}}%
\PRE{%
  \def\frontmatter@abstractheading{\begin{center}\bf Abstract\end{center}}%
  \def\frontmatter@abstractwidth{400pt}%
}%
\makeatother%
\preprint{UCI--TR--2015--02}

\title{ \PRE{\vspace*{1.5in}}%
Long-Lived Sleptons at the LHC and a 100\,TeV Proton Collider%
\PRE{\vspace*{0.3in}}}
\author{Jonathan L.~Feng}
\affiliation{Department of Physics and Astronomy, University of
California, Irvine, CA 92697, USA%
\PRE{\vspace*{.2in}}}
\author{Sho Iwamoto}
\author{Yael Shadmi}
\author{Shlomit Tarem\PRE{\vspace*{.3in}}}%
\affiliation{Physics Department, Technion---Israel Institute of
  Technology, Haifa 32000, Israel%
\PRE{\vspace*{.5in}}}

\begin{abstract}
We study the prospects for long-lived charged particle (LLCP) searches
at current and future LHC runs and at a $100\TeV$ $pp$ collider, using
Drell--Yan slepton pair production as an example. Because momentum
measurements become more challenging for very energetic particles, we
carefully treat the expected momentum resolution.  At the same time, a
novel feature of $100\TeV$ collisions is the significant
energy loss of energetic muons in the calorimeter.  
We use this to help discriminate
between muons and LLCPs.  We find that the $14\TeV$ LHC with an
integrated luminosity of $3\iab$ can probe LLCP slepton masses up to
$1.2\TeV$, and a $100\TeV$ $pp$ collider with $3\iab$ can probe LLCP
slepton masses up to $4\TeV$, using time-of-flight measurements. These
searches will have striking implications for dark matter, with the LHC
definitively testing the possibility of slepton--neutralino
co-annihilating WIMP dark matter, and with the LHC and future hadron
colliders having a strong potential for discovering LLCPs in models
with superWIMP dark matter.
\end{abstract}

\pacs{\tt 12.60.Jv, 95.35.+d, 13.85.-t}

\maketitle

\section{Introduction}\label{sec:intro}

Many extensions of the standard model (SM) predict long-lived charged
particles (LLCPs) that are stable on collider-detector timescales.
Such particles present new challenges for collider experiments,
requiring novel methods for triggering, reconstruction, and detection.
At the same time, their discovery would be extremely exciting, with
profound implications for both particle physics and cosmology. In
addition, LLCPs would provide nearly background-free handles to
discover heavier new particles, if these exist.  For these reasons,
LLCP searches have attracted great interest in recent years,
culminating in new limits on LLCP masses from experiments at the 7
and 8 TeV
LHC~\cite{Chatrchyan:2013oca,ATLAS:2014fka,Khachatryan:2015lla}.

In this paper, we investigate the capabilities of current and future
high luminosity runs of the LHC for discovering LLCPs, as well as the
potential of a future $100\TeV$ hadron collider for LLCP searches.
Because of the unique methods required for their detection, LLCP
searches provide an interesting testing ground for future colliders
and detectors.  In addition, LLCP cosmology and its implications for
future colliders are worth considering.  Cosmology is well-known to
provide constraints that are complementary to conventional particle
physics bounds.  For example, requiring that thermal relic neutralinos
not overclose the Universe implies an {\sl upper} bound on neutralino
masses.  The possibility of completely probing the viable thermal
relic neutralino dark matter (DM) parameter space is therefore useful
input to setting a target center-of-mass energy for future $pp$
colliders~\cite{Gershtein:2013iqa,Low:2014cba,Acharya:2014pua,Gori:2014oua}.
LLCPs may also play key roles in cosmology; for example, they may
decay to DM particles and thereby affect the DM relic abundance.  Here
we determine the implications of cosmological scenarios with LLCPs for
future collider energies and detector design.

We will concentrate on a worst-case scenario, in which the only new
particle within reach is a non-colored LLCP, which we will take to be
a slepton.  Our results are thus based on Drell--Yan slepton pair
production and can be trivially generalized to pair production of
LLCPs with different quantum numbers.  Furthermore, these results are
very robust and do not depend on the assumption of supersymmetry. 

At the same time, supersymmetry provides at least two well-motivated
frameworks for LLCPs. One is gauge-mediated supersymmetry breaking, in
which the lightest supersymmetric particle (LSP) is the gravitino, and
the next-to-lightest supersymmetric particle (NLSP) is a charged
slepton~\cite{Dine:1994vc,Dine:1995ag,Feng:1997zr}.  The reach of the
$100\TeV$ collider for first generation squarks and gluinos has been
estimated to be 10--$15\TeV$~\cite{Cohen:2013xda}.  As we discuss
below, if these particles are beyond reach, the supersymmetry scale
must be high, with the gravitino mass $\gsim{\rm MeV}$ and quite
possibly much higher than that and in the GeV to TeV range.  This
entire range of gravitino masses generically results in a long-lived
slepton NLSP. From the point of view of cosmology, this scenario
provides a realization of superweakly-interacting massive particle
(superWIMP) DM, with metastable sleptons decaying to gravitinos, which
form superWIMP DM~\cite{Feng:2003xh,Feng:2003uy}.

A second framework of interest is the slepton--neutralino
co-annihilation scenario, in which a small slepton--neutralino mass
difference is motivated by DM~\cite{Griest:1990kh,Ellis:1998kh}.  Here
DM is the neutralino LSP, and its relic abundance is diluted through
co-annihilations with a quasi-degenerate slepton.  Slepton decay to
the LSP is thus phase-space suppressed. The correct relic abundance is
obtained for slepton masses $\lsim 600\GeV$.  We will find that, in
agreement with Refs.~\cite{Konishi:2013gda,Desai:2014uha}, the entire
cosmologically-motivated mass range can be probed by the $14\TeV$ LHC.

Non-colored LLCPs interact in the detector much like muons.  Thus, the
main challenge in their discovery is distinguishing them from muons.
ATLAS and CMS rely both on differences in the energy loss ($\dd E/\dd
x$) of LLCPs and muons in the inner detectors, and on time-of-flight
(ToF) measurements in the muon detectors. In this study, we will only
consider the latter, essentially extrapolating from what has been done
at the LHC\footnote{%
  The LHC reach for long-lived slepton was also studied
  in~\cite{Heisig:2011dr,Heisig:2012zq}, selecting sleptons with
  speeds $0.6<\beta<0.8$ to discriminate them from muons.
  Here we select a wider range of slepton $\beta$, based on
  current ATLAS searches.
}. At a $100\TeV$ collider, however, we have a qualitatively
new handle at our disposal, since energetic muons lose energy through
radiative processes, i.e., bremsstrahlung, electron pair-production,
and photo-nuclear interactions~\cite{Groom:2001kq}, in addition to
ionization.  In contrast, the radiative energy loss would be
negligible for a heavy LLCP.
We therefore cut on the energy measured in the calorimeter along 
the track of the candidate, to reduce the number of background muons.

As noted above, a $100\TeV$ collider may provide a definitive test of
(stable) supersymmetric WIMP
DM~\cite{Gershtein:2013iqa,Low:2014cba,Acharya:2014pua,Gori:2014oua}.
We will find that the superWIMP DM scenario is harder to probe
exhaustively, since the DM relic abundance does not provide a strict
upper bound on the slepton mass---increasing the slepton mass can in
principle be compensated by decreasing the gravitino mass.  In the
framework we consider here, the lower bound on the slepton lifetime
$\gsim$~nsec, implies a {\sl model-independent} upper bound on the
slepton mass around $40\TeV$, which is, of course, beyond the reach of
any foreseeable collider.  Still, as we will see, the $100\TeV$
collider with $3\iab$ could probe sleptons with masses up to 3.2 to
4.0 TeV, depending on the left--right composition of the sleptons.
The testable mass range therefore includes a wide range of
cosmologically-allowed models, including the interesting region of
superWIMP models in which late slepton decays may have measurable
effects on big bang nucleosynthesis (BBN) or the cosmic microwave
background (CMB).  As noted above, the worst-case scenario we
consider, with colored superpartners beyond reach, implies a high
supersymmetry-breaking scale, which is precisely the relevant region
for the $100\TeV$ collider LLCP searches.

This paper is organized as follows.  In~\secref{target} we review the
two long-lived slepton scenarios discussed above and summarize the
relevant mass ranges.  In~\secref{coll}, we discuss LLCP collider
searches, starting with a detailed description of our analysis of the
$14\TeV$ LHC in~\secref{14}, and providing an overview of our Monte
Carlo simulation.  We then go on to discuss the $100\TeV$ collider
in~\secref{100}, where we review the proposed detector, discuss novel
features at these extreme energy scales, and study the prospects for
LLCP searches at $100\TeV$.  The results are discussed
in~\secref{disc}.  We conclude with a collection of the results and
some remarks in~\secref{concl}.  Details of the Monte Carlo
simulations are collected in the Appendix.

\section{Target Mass Ranges from Cosmology}\label{sec:target}

The search for LLCPs is important independent of any theoretical
framework, and the searches described in the following sections are in
fact model-independent.  At the same time, it is useful to have some
scenarios with target mass ranges in mind to motivate the searches.
In this section, we highlight two cosmological scenarios that point to
particularly interesting mass ranges for long-lived sleptons.

\subsection{Slepton SuperWIMP Scenarios}

Sleptons may be long-lived, because their decays are mediated by very
weak interactions.  Perhaps the most generic possibility is the
superWIMP scenario~\cite{Feng:2003xh,Feng:2003uy} with slepton NLSPs
that decay to gravitino LSPs, in which the decays are suppressed by
the weakness of gravity. The gravitinos then comprise part, or all, of
DM.

The width for the decay of a slepton to a gravitino
is~\cite{Feng:2004mt}
\begin{equation}
 \Gamma(\tilde{l} \to l \tilde{G}) =\frac{1}{48\pi M_*^2}
 \frac{m_{\tilde{l}}^5}{m_{\tilde{G}}^2}
 \left[1 -\frac{m_{\tilde{G}}^2}{m_{\tilde{l}}^2} \right]^4 \ ,
\label{sfermionwidth}
\end{equation}
where $M_* \simeq 2.4 \times 10^{18}\GeV$ is the reduced Planck mass,
assuming the lepton mass is negligible.  When the gravitino is much
lighter than the slepton, the slepton lifetime is
\begin{equation}
 \tau(\tilde{l} \to l \tilde{G}) \simeq 5.7\times 10^{-7} \sec
 \left(\frac{\TeV}{m_{\tilde{l}}}\right)^5\, 
\left(\frac{m_{\tilde{G}}}{\MeV}\right)^2 \ ,
\label{sfermiontau}
\end{equation}
and for $m_{\tilde{l}} \sim {\rm TeV}$ and $m_{\tilde{G}} \agt
{\rm MeV}$, the slepton is effectively stable in collider experiments.

In superWIMP scenarios, the NLSP first freezes out with relic density
given approximately by~\cite{Bernstein:1985th,Scherrer:1985zt}
\begin{eqnarray}
\Omega_{\text{NLSP}}^{\text{th}} h^2 \approx
\frac{1.1\times 10^9 \, x_F \GeV^{-1}}
{\sqrt{g_*} \, \mplanck \, \langle\sigma v\rangle}
\approx 0.2 \Biggl[\frac{15}{\sqrt{g_*}}\Biggr]
\Biggl[\frac{x_f}{30}\Biggr]
\Biggl[\frac{10^{19}\GeV}{\mplanck}\Biggr]
\Biggl[\frac{10^{-9}\GeV^{-2}}{\langle\sigma v\rangle}\Biggr]
\, ,
\label{relic}
\end{eqnarray}
where $g_*$ is the effective number of massless degrees of freedom at
freeze out, $x_f \equiv m_{\text{NLSP}}/T_f \approx 25$ is the NLSP
mass divided by the freeze out temperature $T_f$, $\mplanck \simeq 1.2
\times 10^{19}\GeV$ is the (unreduced) Planck mass, and $\langle\sigma
v\rangle$ is the thermally-averaged NLSP annihilation cross section.
Let us assume that the NLSPs are right-handed sleptons $\tilde{l}\s{R}$,
and the number of slepton generations among $\tilde e\s{R},
\tilde\mu\s{R}$ and $\tilde \tau\s{R}$ that are degenerate and
long-lived is $N\s{gen;LL}$, where $1\le N\s{gen;LL}\le 3$.  (It is
not difficult to generalize this to scenarios with left-handed slepton
NLSPs.) The dominant annihilation channels are typically $\tilde{l}
\tilde{l}^* \to \gamma \gamma, \gamma Z, ZZ$ through slepton exchange
and $\tilde{l} \tilde{l} \to l l$ through Bino exchange.  For
right-handed sleptons, the thermally-averaged cross section near
threshold is approximately~\cite{Asaka:2000zh}
\begin{eqnarray}\label{sigmav}
\langle \sigma v \rangle 
&\approx& \frac{4\pi \alpha^2}{m_{\tilde{l}\s R}^2} +
 \frac{16\pi\alpha^2m_{\tilde{B}}^2}
{\cos^4 \theta_W (m_{\tilde{l}\s R}^2 + m_{\tilde{B}}^2 )^2}
\equiv C_{\tilde{B}} \frac{4\pi \alpha^2}{m_{\tilde{l}\s R}^2} \ ,
\end{eqnarray}
where $m_{\tilde{B}}$ and $m_{\tilde{l}\s R}$ are the Bino and slepton
masses, respectively, and $C_{\tilde{B}}$ is 1 for infinitely heavy
Binos and increases monotonically to $C_{\tilde{B}} \simeq 2.7$ as the
Bino mass decreases from infinity to near the slepton mass.

When the slepton decays to the gravitino $\tilde{l}\s{R} \to l
\tilde{G}$, the gravitino inherits the relic density
\begin{eqnarray}
\label{omega}
\Omega_{\gravitino} h^2
= \frac{m_{\gravitino}}
{m_{\tilde{l}\s R}}\Omega_{\text{NLSP}}^{\text{th}} h^2
\end{eqnarray}
from each slepton.
Combining Eqs.~(\ref{relic}), (\ref{sigmav}), and (\ref{omega}), we
find that, numerically, the gravitino relic density is
\begin{eqnarray}
\label{combined}
\Omega_{\gravitino} h^2 = 
N\s{gen;LL} \cdot 0.12 \, \frac{ m_{\tilde{l}\s R} m_{\tilde{G}} } {M^2} \ ,
\end{eqnarray}
where $M$ varies from $650\GeV$ to $1.0\TeV$ for Bino masses varying
from $m_{\tilde{B}} = \infty$ to $m_{\tilde{l}\s R}$.  For
$m_{\tilde{l}\s R} \sim m_{\tilde{G}}$, the relic abundance is saturated
when both masses are around $600\GeV$. If, on the other hand, the
gravitino is much lighter than the slepton, the constraint that the
slepton be long-lived, with, say, 
$\tau\s{NLSP}\geq 10^{-6}\sec$~\footnote{A somewhat smaller value, say, $5\times
  10^{-7}\sec$ is probably safe, too, but shorter lifetimes will lead
  to smaller efficiencies, as some of the sleptons may decay in the
  detectors.}, implies
\begin{equation}
\frac{m_{\tilde{l}\s R}}{\rm TeV} \lsim
\left(\frac{m_{\tilde{G}}}{\rm MeV}\right)^{2/5}\,.
\end{equation}
The DM abundance then provides a {\sl model-independent} upper limit
on the slepton mass in this scenario,
\begin{equation} 
m_{\tilde{l}\s R} \lsim 40\TeV\,, 
\end{equation}
which is beyond the reach of any foreseeable experiment.

\begin{figure}[tp]
 \includegraphics[width=0.6\textwidth]{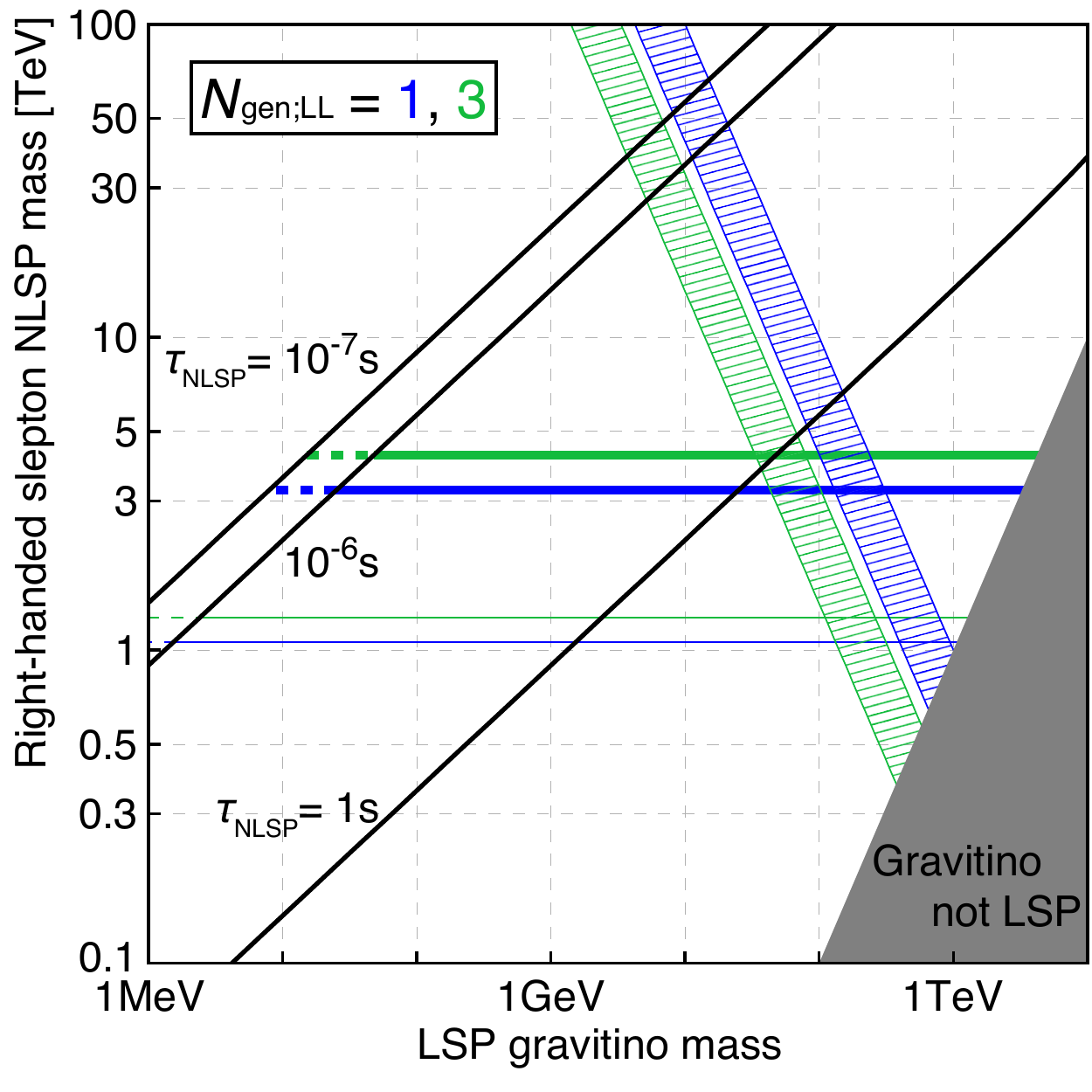}
\caption{An overview of the parameter space in superWIMP scenarios.
  The black lines illustrate the lifetime of the NLSP slepton
  $\tau\s{NLSP}=10^{-7}$, $10^{-6}$ and $1\sec$.  In the blue (green)
  hatched region, gravitinos saturate the DM relic density if
  $N\s{gen;LL}=1$ (3), i.e., one (three) right-handed slepton is
  long-lived.  The upper and lower edges of the regions correspond to
  $m_{\tilde B}\sim m_{\tilde l\s R}$ and $m_{\tilde B}\gg m_{\tilde
    l\s R}$, respectively. The horizontal lines are the expected reach
  of right-handed long-lived slepton searches: the thinner lines are
  the expected exclusion limits at the $14\TeV$ LHC, and the thicker
  lines are the expected exclusion limits at a $100\TeV$
  collider, both with an integrated luminosity of $3\iab$.}
\label{fig:superwimpparameters}
\end{figure}

We display the results above in~\figref{superwimpparameters}, in the
slepton--gravitino mass plane.  In the blue (green) hatched regions,
gravitinos from late slepton decays saturate the DM abundance for one
(three degenerate) long-lived sleptons, depending on the Bino mass.
The upper (lower) edges correspond to $m_{\tilde B}\sim m_{\tilde l\s
  R}$ ($m_{\tilde B}\gg m_{\tilde l\s R}$), i.e., $C_{\tilde B}=2.7$
(1.0) in \eqref{sigmav}, and $M/\sqrt{N\s{gen;LL}}=1\TeV$ ($650\GeV$)
in~\eqref{combined}.  The region of the plane above the upper edge is
excluded by DM overabundance.  

The black lines correspond to different slepton lifetimes.  Above the
10$^{-7}$--$10^{-6}\sec$ lines, some sleptons decay inside the
detector, and the efficiency of the searches described here
deteriorates.  (Of course, this may lead to spectacular signals in
other channels.)  Also shown is the $\tau\s{NLSP}\sim{\rm sec}$ line,
below which BBN and CMB constraints become relevant. For such long
decay times, the SM particles produced in slepton decays are not
quickly thermalized, and they may destroy light elements or modify the
black body spectrum of the CMB~\cite{Feng:2003xh,Feng:2003uy,%
  Feng:2004zu,Feng:2004mt,Kawasaki:2008qe,Bailly:2008yy}.  These
effects may be in conflict with the successes of standard BBN or
observations of the CMB, excluding some late decay scenarios. On the
other hand, in some cases, the late decays may alleviate discrepancies
between the predictions of standard BBN and the observed abundances,
particularly of $^7$Li and $^6$Li.  In any case, it is clear that LLCP
collider probes of the region of parameter space with slepton
lifetimes longer than 1 second may have particularly interesting
implications for the early Universe.

Finally, we also show in this figure the main results of the analysis
of \secsref{14}{100}, namely, the projected reach of the $14\TeV$ LHC
and $100\TeV$ $pp$ collider.  These are given by the horizontal lines,
the thinner for the $14\TeV$ LHC and the thicker for the $100\TeV$
collider, where again, the blue (green) line corresponds to one (three
degenerate) right-handed sleptons. We see that collider searches can
probe a significant portion of the allowed parameter space, including
most of the superWIMP parameter space with lifetimes longer than a
second, which, as explained above, is especially interesting.

In fact, the region which could be probed by a $100\TeV$ collider is
also well-motivated by more theoretical considerations.  Recall that
we assume here that squark and gluino masses are above $10\TeV$ and
beyond the reach of a $100\TeV$ collider.  In gauge-mediation, these
masses are roughly given by
\begin{equation}
10^{-2}\, \frac{F_{{\rm GMSB}}}{M_{{\rm mess}}} \,,
\end{equation}
with $M_{{\rm mess}}^2 > F_{{\rm GMSB}}$.  Thus, both the messenger
scale $M_{{\rm mess}}$ and the supersymmetry breaking $F_{{\rm GMSB}}$
are pushed to high values.  The gravitino mass is given by
\begin{equation}
  m_{\tilde{G}} \sim \frac{F_0}{\mplanck} \equiv c_{{\rm grav}}
  \frac{F_{{\rm GMSB}}}{\mplanck} \, ,
\end{equation}
where $F_0$ is the dominant supersymmetry-breaking $F$-term.  The
number $c_{{\rm grav}}$ depends on the details of the supersymmetry
breaking sector; the most concrete, calculable models predict $c_{{\rm
    grav}} \gg 1$.  Combining these, we see that $m_{\tilde{G}} \agt
1\MeV$, with values of $1\text{--}100\GeV$ perhaps even more
plausible, and so TeV-mass sleptons are necessarily long-lived on
collider-detector timescales.

Let us briefly discuss the limit obtained at the $8\TeV$ LHC, which is
not shown in the figure. Assuming only Drell--Yan direct pair
production of a single generation slepton ($N\s{gen;LL}=1$), the CMS
(ATLAS) Collaboration excludes long-lived sleptons with masses
$m<346\,(286)\GeV$~\cite{Chatrchyan:2013oca,ATLAS:2014fka}, which does
not exclude any of the region suggested in the SuperWIMP scenario (the
hatched region). For $N\s{gen;LL}=3$, the ATLAS Collaboration excludes
$m<337\GeV$, and the CMS analysis excludes $m\lesssim440\GeV$, which
slightly overlaps the cosmologically-favored region.

\subsection{Slepton--Neutralino Co-Annihilation}

Sleptons may be long-lived because their decay rate is phase-space
suppressed.  Perhaps the best motivation for such phase-space
suppression is the slepton--neutralino co-annihilation scenario, in
which neutralinos freeze out and are DM, and their thermal
relic density is reduced to viable levels through co-annihilation with
highly degenerate sleptons.

This has recently been explored in detail in
Refs.~\cite{Konishi:2013gda,Desai:2014uha} in the CMSSM framework,
where there is a cosmologically-preferred stau--neutralino
co-annihilation region of parameter space, but the resulting ranges of
neutralino and stau masses hold more generally, since they are driven
by the DM relic abundance.  For stau--neutralino splittings less than
about $1\GeV$, the staus are long-lived at colliders, and the correct
relic density can be obtained for gaugino masses $M_{1/2} \sim
800\text{--}1400\un{GeV}$, where the exact value depends on
$\tan\beta$ and the $A$-parameter that determines the left--right stau
mixing.  This scenario therefore motivates stau masses
\begin{equation}
m_{\tilde{\tau}} \simeq m_{\chi} \simeq 0.42 M_{1/2}
\approx 350\text{--}600\GeV \ .
\end{equation}
This range is just being probed by current bounds.  The upper bound is
achieved for exactly degenerate staus and neutralinos, where the
co-annihilation effect is maximized, and so this is a hard upper bound
in this scenario: heavier staus will necessarily overclose the
Universe.

\section{LLCP Collider Searches}\label{sec:coll}

In collider experiments, metastable sleptons, or more generally
non-colored LLCPs, interact with the detectors much like muons.  An
LLCP passes through the detector, leaving a charged track from
ionization energy loss, with small energy deposits in the
calorimeters.  Therefore, the main background is muons, and the only
difference between a hypothetical LLCP and a muon is the (assumed)
large mass of the former.
Because of this large mass, LLCPs would typically be produced with
a smaller speed $\beta$.
This speed can be measured using the ToF to the outer detectors,
or the ionization energy loss, $\dd E/\dd x$, which depends on
$\beta\gamma$, with $\gamma=(1-\beta^2)^{-1/2}$. 
The ATLAS and CMS collaborations have
used both of these methods at the LHC with $\sqrt s=7\text{--}8\TeV$, but here
we will only consider ToF measurements, for which the specifics of the
detector are less relevant.

At very high energies, the LLCP mass leads to an additional
qualitative difference between LLCPs and muons: while TeV-energy
muons lose significant energy through radiative processes, LLCPs do not.
This can provide a useful handle for discriminating LLCPs from muons
at future high energy colliders.

As noted above, we consider a worst-case scenario in which the only
new particles produced are slepton LLCPs.  The signal is, then,
Drell--Yan slepton pair production, and we will consider three
different slepton types: purely left-handed sleptons, which we denote
$\tilde{e}\s L$, purely right-handed sleptons $\tilde{e}\s R$, and
left--right mixed sleptons, which we denote $\tilde{\tau}_1$. Note
that slepton flavor does not matter here, since the slepton does not
decay in the detector\footnote{If other superpartners are also within
  reach, production of these particles would lead to much higher reach
  in the LLCP mass because such events typically include at least two
  LLCPs with accompanying visible particles.  Note that the $\beta$
  distributions of such LLCPs tend to be stiffer.}.

In the following we will study the prospects for slepton detection at
a 100 TeV collider, as compared to the 14 TeV LHC.  There is no
concrete design, at this point, of the detectors that will be deployed
at a $100\TeV$ collider.  Furthermore, detector techniques are
expected to improve before such a design is made. We therefore make
several simplifying assumptions. The main one is that the detector
will measure the momenta of high-momentum particles produced at $\sqrt
s=100\TeV$ as well as the LHC detectors perform for particles with momenta up
to 1 TeV. A second assumption is that new advances will allow good
resolution at high pile-up, or that the collider will not run at
luminosities so high that the pile-up will prevent good
reconstruction.

The uncertainty regarding the detector performance far outweighs
the effects of systematic uncertainties, on the
order of 10--20\%, that were assigned in the LHC Run 1
searches~\cite{Chatrchyan:2013oca,ATLAS:2014fka}.
Thus, for meaningful comparison of the 14 TeV and 100 TeV searches,
we do not consider systematic uncertainties in this work.

\subsection{LLCP Searches at the 14\,TeV LHC}\label{sec:14}

\subsubsection{Monte Carlo Simulation}

We use the Snowmass background set for $14\TeV$ $pp$
colliders~\cite{Anderson:2013kxz,Avetisyan:2013onh,Avetisyan:2013dta},
which is briefly described in the Appendix, to estimate SM background.
We generate our signal events, slepton Drell--Yan pair production,
with the same tools used to generate the background set. The pair
production is calculated at tree-level using
\software{MadGraph5\_aMC@NLO}~\cite{MG5AMC}, with showering and
hadronization performed by~\software{Pythia\,6}~\cite{Pythia6.4} with
the~\software{Pythia-PGS} interface.  For the detector simulation we
use~\software{Delphes} tuned by the Snowmass Collaboration based on
\software{Delphes\,3.0.9}~\cite{deFavereau:2013fsa,Cacciari:2011ma,Cacciari:2005hq}.
The momentum resolution of muons is assumed to be $\Delta\PT=0.05\PT$
for $\PT>200\GeV$.  Pileup is not considered.

Because the ToF measurement is used to distinguish sleptons from
muons, its resolution is carefully treated.  At the ATLAS detector,
the resolution of ToF is reported as $2.5\%$~\cite{ATLAS:2014fka}.
This is the value we use for the slepton speed measurement.  Thus, the
slepton speed is smeared according to
\begin{equation}\label{beta1}
 \mathop{\mathrm{PDF}}(\hat\beta^{-1})_{\tilde l} = \Normal(\beta^{-1}, 0.025),
\end{equation}
where $\hat\beta$ is the smeared slepton speed, and
$\Normal(\mu,\sigma)$ is the Normal distribution with mean $\mu$ and
dispersion $\sigma$.  For muons, however, this distribution is
inaccurate, because the dominant background comes from the tail of the
distribution.  We therefore use a more detailed distribution for the
muons' $\hat\beta$,
\begin{equation}\label{beta2}
 \mathop{\mathrm{PDF}}(\hat\beta)_\mu
  = 0.832\cdot \Normal(1, 0.022) + 
0.162 \cdot \Normal(1, 0.050) + 0.00534 \cdot \Normal(1, 0.116)\,,
\end{equation}
which is obtained by fitting the measured $\beta$ distribution
at the ATLAS experiment (Fig.~1 of Ref.~\cite{ATLAS:2014fka}).

After object identification performed by \software{Delphes}, all
objects with $\PT<30\GeV$ are dropped, and muon pairs are removed if
their invariant masses satisfy $|m_{\mu\mu}-m_Z|<5\GeV$.  The
remaining muons are tagged as LLCPs if they satisfy the following
conditions:
\begin{itemize}
 \item $\PT>100\GeV$ and $|\eta|<2.4$,
 \item $\Delta R>0.5$ from the nearest reconstructed object (with
   $\PT>30\GeV$),
 \item $0.3<\hat\beta<0.95$,
\end{itemize}
where $\hat\beta$ is the smeared speed as defined above.

The accurate measurement of the speed $\beta$ is a result of quality
requirements made on the reconstructed tracks and timing measurements.
Following the results of the ATLAS selection, we assign quality
selection efficiencies of $\epsilon_\mu = 0.5$ for identifying a fake
LLCP (muon), and $\epsilon_{\tilde l}=0.6$ for a true LLCP
(slepton)\footnote{The efficiency for fake LLCP identification is
  worse than for true LLCP identification, because a poorly measured
  $\beta$ is uncorrelated between sub-detectors, and may be correlated
  with a poor-quality track.}.

We select events with two LLCP candidates.  If the event has more than
two LLCP candidates, the two with the highest $\PT$'s are used.  For
each LLCP, we calculate the reconstructed mass
\begin{equation}
 \hat m = \frac{\PT\cosh\eta}{\hat\beta\hat\gamma} \, ,
\end{equation}
where $\hat\gamma=(1-\hat\beta^2)^{-1/2}$.

We define eight signal regions (SRs): SR300, SR400, $\ldots$, SR1000,
where SR$x$ requires both of the LLCPs to have $\hat m > x\GeV$.  For
each signal region, the expected 95\% confidence level (CL) upper
limit on the number of events, $N\s{UL}$, is calculated with the ${\rm
  CL}_s$ method~\cite{Read:2002hq}.  Based on $N\s{UL}$, the
corresponding upper limit on the signal cross section, $\sigma\s{UL}$,
is calculated for different LLCP scenarios.  Because of the inclusive
SR definition, the lowest $\sigma\s{UL}$ gives the limit on the
scenario.  Statistical uncertainties are considered, but systematic
uncertainties are not included in this analysis.

\subsubsection{Results}

The LLCP selection flow is shown in \tableref{selectionflow-14} for
several LLCP masses, together with the total cross sections and the
cross sections for events with one and two tagged LLCPs.  Note that
the signal is calculated at LO, while the background is calculated at
NLO.  The efficiency factors $\epsilon_\mu$ and $\epsilon_{\tilde l}$
are not imposed in this table for simplicity.

In \tableref{SRs-14}, we show the separate contributions in each of
the signal regions, with the different efficiencies for sleptons and
fake LLCPs included.  We also display $N\s{UL}$ for integrated luminosities
of~$\int\mathcal L=0.1$, 0.3, and $3\iab$.  Tighter SRs are mostly
background free, and result in $N\s{UL}\simeq3.0$ because of the
statistical uncertainty due to the Poisson distribution.

\begin{table}[t]
\catcode`?=\active \def?{\phantom{0}} \catcode`@=\active \def@{\phantom{.}}
 \caption{LLCP selection flow and cross section of events in the
   $14\TeV$ LHC analysis, for 
   $\tilde e\s L$ pair-production ($N\s{gen;LL}=1$).
   The efficiency factors $\epsilon_\mu$
   and $\epsilon_{\tilde l}$ are not included.   SM background
   (BKG) is calculated with NLO cross sections, while signal cross
   sections are based on Drell--Yan production at tree-level.  We show
   the number of LLCP candidates for LLCP selection flow, and the number
   of events for event cross section.}
 \label{table:selectionflow-14}
\begin{tabular}[t]{lccccc}
 \toprule
                         & \multicolumn{4}{c}{signal ($pp\to\tilde e\s L\tilde e\s L^*$) with $m_{\tilde e\s L}=$}  & SM BKG \\
   &  ~~$400\GeV$~~  & ~~$600\GeV$~~ & ~~$800\GeV$~~ & ~~$1\TeV$~~ & --- \\
 \colrule
 \multicolumn{6}{l}{LLCP selection flow [ab]}\\
 \colrule
 candidates                & $2.31\EE3$ & 359  & 80.5 & 21.9 & ---\\
 + $\PT>100\GeV$, isolated & $2.08\EE3$ & 337  & 76.4 & 20.9 & $1.06\EE8$\\
 + $0.3<\hat\beta<0.95$    & $1.77\EE3$ & 312  & 73.9 & 20.6 & $3.92\EE6$\\
 \colrule
 \multicolumn{6}{l}{Event cross section [ab]}\\
 \colrule
 total cross section       & $1.15\EE3$ & 180  & 40.2 & 10.9 & ---\\
 $N\s{LLCP}=1$             & 320        & 35.8 & 5.55 & 1.10 & $3.92\EE6$ \\
 $N\s{LLCP}=2$             & 727        & 138  & 34.2 & 9.74 & $1.29\EE3$\\
 \botrule
 \end{tabular}
\end{table}

\begin{table}[p]
\catcode`?=\active \def?{\phantom{0}} \catcode`@=\active \def@{\phantom{.}}
 \caption{Contributions to SRs in the $14\TeV$ LHC analysis with
   efficiency factors ($\epsilon_\mu$ and $\epsilon_{\tilde l}$)
   included, and 95\% CL upper limits on the number of events
   ($N\s{UL}$) for integrated luminosities $\int {\cal L} =0.1$, 0.3,
   and $3\iab$, based on $\tilde e\s{L}$ production ($N\s{gen;LL}=1$).
   Statistical uncertainties are considered but systematic
   uncertainties are not included.  In the columns of signal event
   contributions, bold numbers mark the SR which gives the lowest
   ${\rm CL}_s$ in the analysis of $\int\mathcal L=0.3\iab$, and
   contributions less than $0.1\ab$ are displayed as zero.}
 \label{table:SRs-14}
\begin{tabular}[t]{l|cccc|c|ccc}
 \toprule
               & \multicolumn{4}{c|}{signal ($pp\to\tilde e\s L\tilde e\s L^*$)~[ab] with $m_{\tilde e\s L}=$} & SM BKG
               & \multicolumn{3}{c}{$N\s{UL}$ with $\int\mathcal L=$}\\
               & ~~$400\GeV$~~  & ~~$600\GeV$~~ & ~~$800\GeV$~~ & ~~$1\TeV$~~ & [ab] & $0.1\iab$ & $0.3\iab$ & $3\iab$\\
 \colrule
 $N\s{LLCP}=2$ & 262        &  50        &  12        & 3.5       &  323       & --- & --- & --- \\
 SR300         & {\bf259}   &  50        &  12        & 3.5       &  2.8       & 3.4 & 4.0 & 7.5 \\
 SR400         & 74         &  50        &  12        & 3.5       &  0.67      & 3.1 & 3.3 & 4.9 \\
 SR500         & 0.85       & {\bf 47}   &  12        & 3.5       &  0.19      & 3.0 & 3.1 & 3.8 \\
 SR600         &     0      &  13        & {\bf 12}   & 3.5       & $6\EE-2$ & 3.0 & 3.0 & 3.3 \\
 SR700         &     0      & 0.28       &  11        &{\bf 3.5}  & $2\EE-2$ & 3.0 & 3.0 & 3.1 \\
 SR800         &     0      &    0       &  3.1       & 3.4       & $6\EE-3$ & 3.0 & 3.0 & 3.0 \\
 SR900         &     0      &    0       &    0       & 3.0       & $2\EE-3$ & 3.0 & 3.0 & 3.0 \\
 SR1000        &     0      &    0       &    0       & 0.86      & $<10^{-3}$ & 3.0 & 3.0 & 3.0 \\
 \botrule
 \end{tabular}
 \end{table}

\begin{figure}[p]
\includegraphics[width=0.8\textwidth]{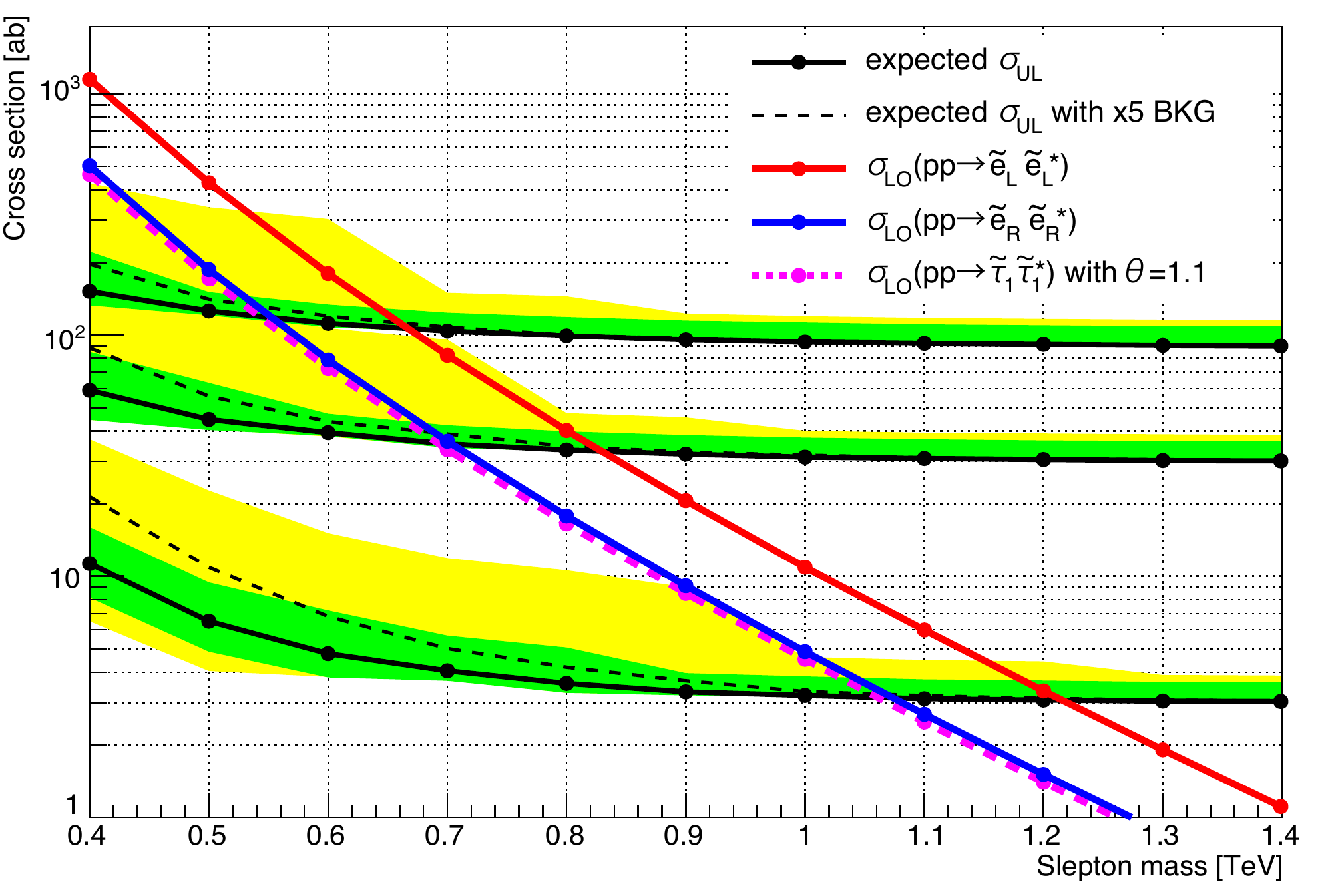}
\caption{Summary plot of the potential reach of LLCP searches at the
  $14\TeV$ LHC.  The thick black lines give the expected upper limits
  $\sigma\s{UL}$ on the signal cross section for integrated
  luminosities $\int \mathcal L=0.1$, 0.3, and $3\iab$ from top to
  bottom.  Green (yellow) bands show the $68\%$ ($95\%$) statistical
  uncertainty regions of $\sigma\s{UL}$.  To estimate the effect of
  systematic uncertainty on the background, the expected
  $\sigma\s{UL}$ assuming a five-times larger background
  is also shown (black dashed lines).  Signal cross sections are
  calculated at tree-level and drawn as red, blue, and magenta solid
  lines, assuming Drell--Yan pair production of $\tilde e\s{L}\tilde
  e\s{L}^*$, $\tilde e\s{R}\tilde e\s{R}^*$, and $\tilde \tau_1\tilde
  \tau_1^*$ with the stau mixing angle $\theta=1.1$, respectively.  }
\label{fig:Limit14}
\end{figure}

The results of this analysis are shown in~\figref{Limit14}.  The upper
bound $\sigma\s{UL}$ (black solid lines) on the signal cross section
is computed for integrated luminosities $\int\mathcal L=0.1$, $0.3$,
and $3\iab$.  The statistical uncertainty is indicated by the green
and yellow bands; the observed limits would fall in the green (yellow)
band with a probability of $68\%$ ($95\%$).  To quantify the effect of
systematic uncertainties on the background, we calculated
$\sigma\s{UL}$ with the background contribution multiplied by five
(black dashed lines).

The signal cross sections are also given by the solid contours.  That
for left- (right-) handed sleptons is drawn by the red (blue) contour.
For the left--right mixed slepton $\tilde\tau_1$, the Drell--Yan
production cross section is maximized in the case where $\tilde\tau_1$
coincides with $\tilde\tau\s{L}$, and minimized for $\theta\simeq 1.1$,
where we define $\tilde\tau_1=\tilde\tau\s L\cos\theta + \tilde\tau\s
R\sin\theta$.  Thus the cross section at $\theta=1.1$ is given by the
solid magenta line, so that the expected reach of mixed slepton LLCP
search lies between the magenta and red lines for any value of
$\theta$.

We see that, for $N\s{gen;LL}=1$, long-lived left-handed
(right-handed) sleptons below $\sim 800\,(700)\GeV$ can be excluded by
Run~2 of the LHC with $\int \mathcal L=0.3\iab$, and below $\sim
1.2\ (1.1)\TeV$ at the high-luminosity LHC (HL-LHC) with $\int
\mathcal L = 3\iab$.  These numbers assume single slepton
production. Left--right mixed sleptons can be excluded for
$m\s{UL}\simeq700\text{--}800\GeV$, depending on the mixing angle, in
Run~2, and for $1.1\text{--}1.2\TeV$ at the HL-LHC.

It is also interesting to consider scenarios with two or more
(nearly-) degenerate sleptons.  For example, if the right-handed
selectron and smuon are degenerate and long-lived, the limits on their
mass would increase to $0.8\,(1.2)\TeV$ with $\int\mathcal
L=0.3\,(3)\iab$.

\subsection{LLCP Searches at a 100\,TeV $pp$ Collider}
\label{sec:100}

\subsubsection{Detector Assumptions and General Considerations}

We now proceed to analyze slepton pair production in a $100\TeV$
collider.  We assume a detector that is roughly like ATLAS or CMS,
with the collision point and the beam pipe surrounded by an inner
detector (ID) for tracking, followed by calorimeters, and with the
muon spectrometer (MS) as the outermost layer.  We utilize only the
region $|\eta|\lesssim2.5$~\footnote{We assume LLCPs are
  produced by large energy transfer, which results in smaller $|\eta|$
  of the LLCPs.}.

The detectors should meet the following two conditions to achieve good
object reconstruction and particle identification.  First, the
calorimeters should be thick and dense enough to stop electrons,
photons, and hadrons, which guarantees good muon observation at the
MS.  Second, the magnetic fields inside the trackers should be large
enough to bend energetic charged particles.  As we see below, the
momentum resolution is determined by the field strength.  LLCPs are
observed as slow muons, and searched for using ToF techniques employed
at Run~1 of the LHC, which we reviewed at the beginning of
\secref{coll}.

\begin{figure}[tp]
 \includegraphics[width=0.573\textwidth]{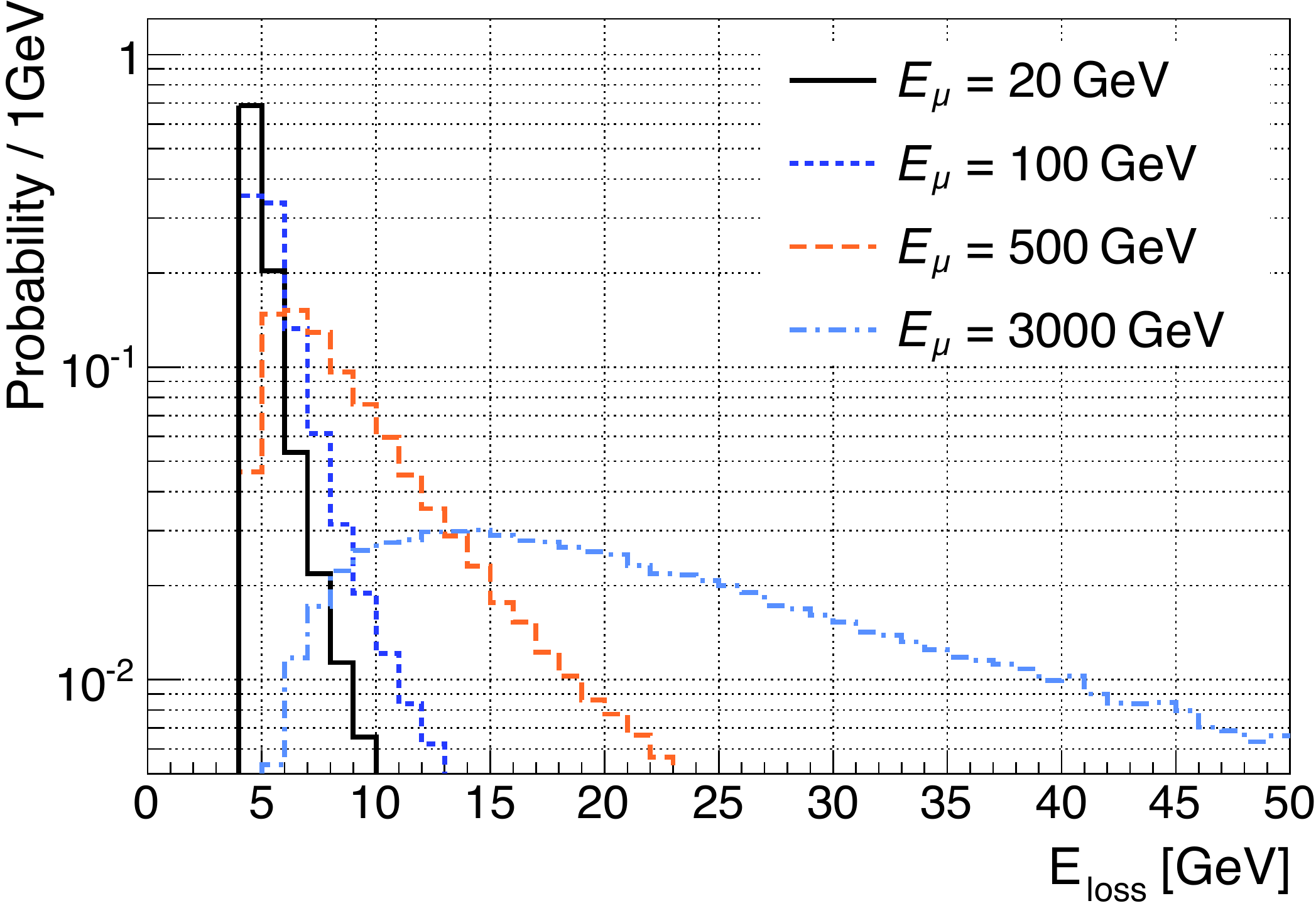}
 \caption{Energy loss of muons in $3\un{m}$ iron. Note that this
   includes the ionization energy loss of 4.8 (6.0) GeV for $E_\mu=20$
   (3000) GeV.}
 \label{fig:muonGeVHist}
\end{figure}

\begin{figure}[tp]
 \includegraphics[width=0.589\textwidth]{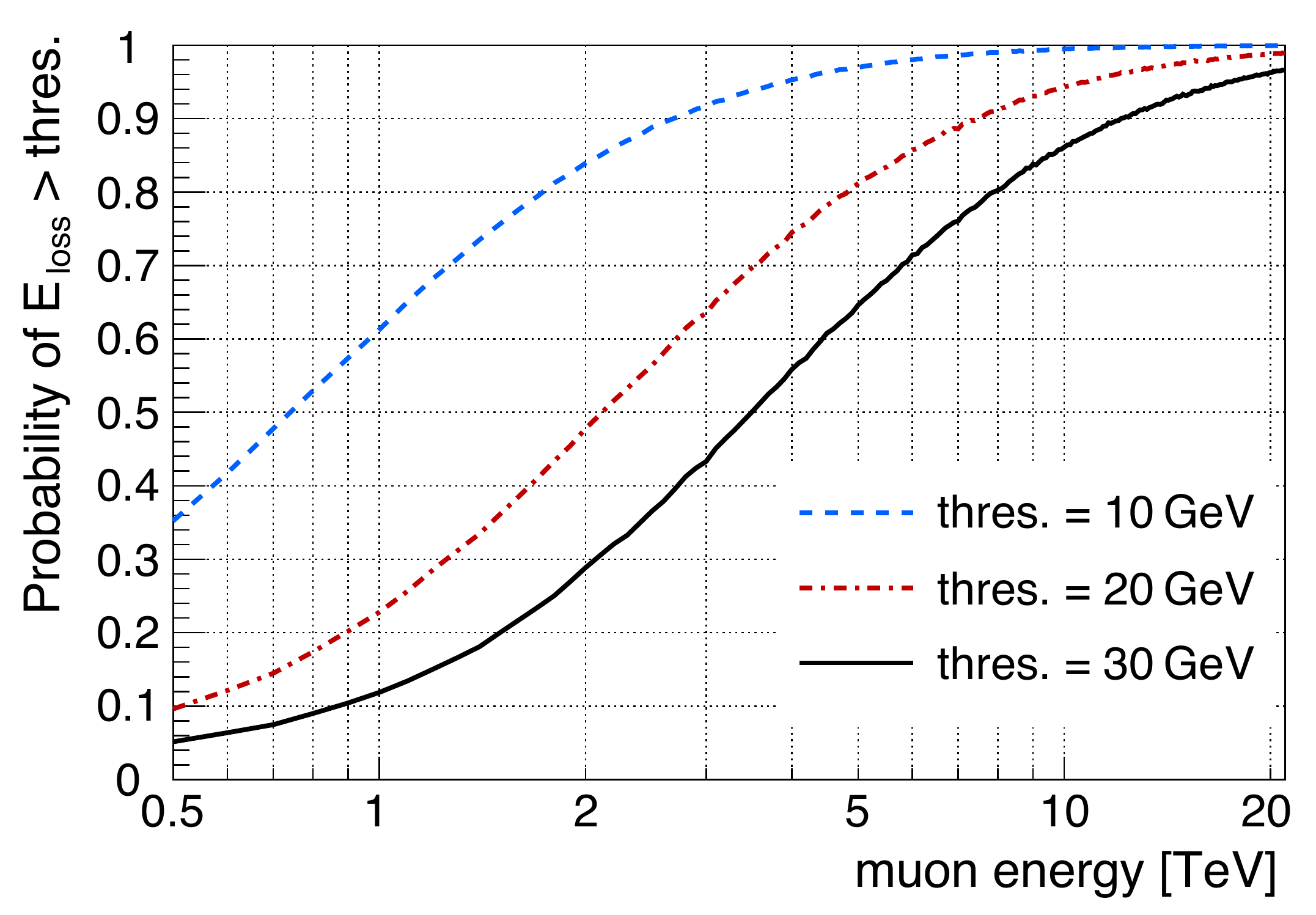}
 \caption{The probability that the energy loss of a muon in a
detector exceeds certain thresholds as a function of the muon energy.
The detector is modeled as $3\un{m}$ of iron.}
 \label{fig:muonProbHist}
\end{figure}

In $100\TeV$ collisions, however, two new features are expected. 
First, muons will deposit more energy in the calorimeters.  At the
LHC, a muon mostly loses its energy by
ionization~\cite{Groom:2001kq}\footnote{See also
  Ref.~\cite[Sec.~32]{PDG2014}}. In iron, the ionization energy loss is
$1.6\text{--}2.0\un{GeV/m}$ for $E_\mu=20\text{--}3000\GeV$.  However,
energetic muons can additionally lose energy through radiative processes.  We
use~\software{Geant\,4.10}~\cite{GEANT4} to estimate this effect.
\Figref{muonGeVHist} illustrates the total energy loss of muons in a
hypothetical $3\un{m}$-thick iron detector.  For $E_\mu\ge500\GeV$,
the energy loss is significant.  The probability that the energy loss
exceeds 10, 20 or $30\GeV$ is shown in~\figref{muonProbHist}.

Second, the $\PT$ resolution of muons is expected to be worse.  In
general, the $\PT$ resolution in trackers can be parametrized as
\begin{equation}\label{momres}
 \Delta\PT = A\oplus B\cdot\PT\oplus C\cdot\PT^2 \, ,
\end{equation}
where the contribution to $A$ is due to muon energy loss before the
tracker, $B$ comes from multiple scattering, and $C$ from the
resolution of position measurements.  For high-$\PT$, the resolution
is therefore dominated by the $\PT^2$ term~\cite{Salvucci:2012np} (see
also Ref.~\cite{Aad:2009wy}),
\begin{align}
 \Delta\PT \approx C\cdot \PT^2 \, .
\end{align}

In our analysis for the $14\TeV$ LHC, the muon $\PT$ resolution was
approximated as $\Delta\PT=0.05\PT$.  This should be supplemented by
the effect of $C$ in analyses of $100\TeV$ collisions.  The value of
$C$ was measured by ATLAS at a very early stage of the $7\TeV$ run to
be $C=0.168(16)\un{TeV^{-1}}$ and $0.417(11)\un{TeV^{-1}}$ for the
barrel region of the MS and ID, respectively~\cite{Salvucci:2012np}.
Since stronger magnetic fields in the tracker, as well as larger
detector dimensions, would improve the momentum resolution, we use
$C=0.1\un{TeV^{-1}}$ in the following analysis\footnote{This
  discussion can be easily understood by approximating the momentum
  measurement as a sagitta measurement.  When a particle of charge $q$
  and momentum $\vc p$ flies a distance $L$ in a magnetic field $B$,
  it has sagitta $s = qL^2B/8p_{\perp}$, where $p_\perp$ is the
  component of $\vc p$ perpendicular to $\vc B$.  Assuming that the
  uncertainty of the sagitta measurement is a constant $\Delta s$, the
  uncertainty of $p_\perp$ is $\Delta p_\perp \simeq 8p_\perp^2\Delta
  s/qL^2B$.}.

\subsubsection{Method}

Our discussion here closely follows the discussion of~\secref{14},
with the two novel aspects being the worse momentum resolution and
muon radiative energy loss discussed above.  As before, slepton pair
production is calculated
using~\software{MadGraph5\_aMC@NLO}~\cite{MG5AMC} at tree-level, with
showering and hadronization performed
by~\software{Pythia\,6}~\cite{Pythia6.4} with
the~\software{Pythia-PGS} interface.  The Snowmass background set for
$100\TeV$ colliders is used for the SM background events, with the
detector assumed to be as described in~\secref{100}.  Pileup is not
considered.

In the Snowmass background set, and thus in the \software{Delphes}
detector simulation in our signal event generation, muon momenta
(for $|\eta|<2.5$ and $\PT>200\GeV$) are smeared according to
$\Delta\PT=0.05\PT$.
We think this is too optimistic for $\PT\gtrsim500\GeV$,
and exploit ``momentum re-smearing'' in object identification.

Object identification and event selection are implemented as follows.
First, after object identification by \software{Delphes}, all objects
with $\PT<100\GeV$ are dropped.  Then, the momenta of the remaining
muons are re-smeared according to the normal distribution
\begin{equation}
 \Normal(\PT, C\cdot \PT^2) \, ,
\end{equation}
where $C=0.1\un{TeV^{-1}}$, and $\PT$ is the momentum after the
\software{Delphes} detector simulation.
After that, muon pairs are removed if their invariant masses satisfy
$|m_{\mu\mu} - m_Z| < 5\GeV$.

For further suppression of background muons, we exploit the muon
radiative energy loss.  Because the background for $m_{\tilde l}\sim 1\TeV$
sleptons under our event selection is from energetic muons with $\PT\gtrsim 500\GeV$, we
can reduce the number of background events by requiring the energy
loss of a candidate LLCP to be below a certain threshold.  
The measured energy loss,  $E\s{loss}$, is the sum of the energy deposits 
along the candidate's trajectory in the calorimeter (corrected for pile-up).
We note that,  while they do not have radiative energy loss, 
true LLCPs have larger energy deposits from ionization
compared to minimum ionizing particles of the same momentum, 
because of their smaller $\beta\gamma$.  
For $m=0.4$ to 3 TeV sleptons, the energy loss in
$3\un{m}$ of iron is estimated as $E\s{loss} = 21.7$, 13.4, and
$9.20\GeV$ for $\beta=0.3$, 0.4, and 0.5, respectively.  Obviously,
the details of detector response and energy resolution will depend on
the actual detector design.  Here we require LLCPs to have
$\hat\beta>0.4$ and $E\s{loss}<30\GeV$, and assume that a true
(slepton) LLCP always satisfies the latter condition\footnote{The
  reduction of signal events due to the tighter $\beta$ cut is
  negligible (less than $2\%$).}.  The $\beta$ resolution is modeled in
the same way as in the $14\TeV$ analysis.  Accordingly, any remaining
muon is tagged as an LLCP if it satisfies the following conditions:
\begin{itemize}
 \item $\PT>500\GeV$ and $|\eta|<2.4$,
 \item $\Delta R>0.5$ from the nearest reconstructed object (with
   $\PT>100\GeV$),
 \item $0.4<\hat\beta<0.95$,
 \item $E\s{loss}<30\GeV$.
\end{itemize}

Events containing two LLCP candidates are selected, and SRs are defined in the
same manner as in \secref{14}, with 16 SRs: SR500, SR600, ..., and
SR2000.  The efficiencies $\epsilon_\mu=0.5$ and $\epsilon_{\tilde
  l}=0.6$ are also imposed.  Statistical uncertainties are considered,
but systematic uncertainties are not included.

  \subsubsection{Results}

The selection flow is presented in~\tableref{selectionflow-100}, with the
cross section broken into different signal regions
in~\tableref{SRs-100}.  The efficiency factors $\epsilon_\mu$ and
$\epsilon_{\tilde l}$, are included in~\tableref{SRs-100} but not
in~\tableref{selectionflow-100}.

\begin{table}[t]
 \caption{LLCP selection flow and cross section of events in the
   $100\TeV$ $pp$ collider analysis, where efficiencies $\epsilon_\mu$
   and $\epsilon_{\tilde l}$ are not included, for the case with
   $\tilde e\s L$ pair-production ($N\s{gen;LL}=1$). The same
   conventions as in \tableref{selectionflow-14} are used.}
 \label{table:selectionflow-100}
\catcode`?=\active \def?{\phantom{0}} \catcode`@=\active \def@{\phantom{.}}
\begin{tabular}[t]{lccccc}
 \toprule
 & \multicolumn{4}{c}{signal ($pp\to\tilde e\s L\tilde e\s L^*$) with $m_{\tilde e\s L}=$}  & SM BKG\\
 & ~~$1\TeV$~~ & ~~$2\TeV$~~ & ~~$3\TeV$~~ & ~~$4\TeV$~~ & --- \\
 \colrule
 \multicolumn{6}{l}{LLCP selection flow [ab]}\\
 \colrule
  candidates               & $2.57\EE3$ & 179  & 31.8 & 8.27 & ---\\
 + $\PT>500\GeV$, isolated & $1.84\EE3$ & 153  & 28.5 & 7.49 & $9.19\EE6$\\
 + $0.4<\hat\beta<0.95$    & $1.23\EE3$ & 121  & 24.6 & 6.83 & $3.41\EE5$\\
 + $E\s{loss}<30\GeV$      &  ---  &  ---        &    ---      &    ---      & $2.78\EE5$\\
 \colrule
 \multicolumn{6}{l}{Event cross section [ab]}\\
 \colrule
 total cross section       & $1.28\EE3$ & 89.5 & 15.9 & 4.14 & ---\\
 $N\s{LLCP}=1$             & 378        & 27.8 & 4.46 & 1.03 & $2.78\EE5$ \\
 $N\s{LLCP}=2$             & 424        & 46.5 & 10.1 & 2.90 & $34.6$ \\
 \botrule
 \end{tabular}
\end{table}

\begin{table}[p]
 \caption{Contributions to SRs in the $100\TeV$ $pp$ collider analysis
   with efficiency factors ($\epsilon_\mu$ and $\epsilon_{\tilde l}$)
   included, and 95\% CL upper limits on the number of events
   ($N\s{UL}$) for integrated luminosities $\int\mathcal L=0.3$, 1,
   and $3\iab$, based on $\tilde e\s L$ pair-production ($N\s{gen;LL}=1$).
   Not all SRs are shown.  Statistical uncertainties
   are considered, but systematic uncertainties are not included. Bold
   numbers mark the SRs that give the lowest ${\rm CL}_s$ in the
   analysis of $\int\mathcal L=1\iab$.}
 \label{table:SRs-100} 
\catcode`?=\active \def?{\phantom{0}} \catcode`@=\active \def@{\phantom{.}}
\begin{tabular}[t]{l|cccc|c|ccc}
 \toprule
 & \multicolumn{4}{c|}{signal ($pp\to\tilde e\s L\tilde e\s L^*$) [ab] with $m_{\tilde e\s L}=$} & SM BKG
               & \multicolumn{3}{c}{$N\s{UL}$ with $\int\mathcal L=$}\\
               & ~~$1\TeV$~~&~~$2\TeV$~~ & ~~$3\TeV$~~&~~$4\TeV$~~ & [ab] & $0.3\iab$ & $1\iab$ & $3\iab$\\
 \colrule
 $N\s{LLCP}=2$ &     153    &     17     &     3.6    &     1.0    &     8.7    & --- & --- & --- \\
 SR500         &     152    &     17     &     3.6    &     1.0    &     1.6    & 3.6 & 4.1 & 6.2 \\
 SR700         &{\bf 146}   &     17     &     3.6    &     1.0    &    0.61    & 3.3 & 3.5 & 4.7 \\
 SR1000        &      43    &{\bf 16}    &     3.5    &     1.0    &    0.14    & 3.1 & 3.1 & 3.6 \\
 SR1200        &     4.0    &     16     &{\bf 3.5}   &{\bf 1.0}   &    0.06    & 3.0 & 3.1 & 3.3 \\
 SR1500        &    0.33    &     13     &     3.3    &    0.97    &    0.03    & 3.0 & 3.0 & 3.1 \\
 SR1700        &    0.10    &     10     &     3.2    &    0.94    &   0.007    & 3.0 & 3.0 & 3.0 \\
 SR2000        &     0      &     4.4    &     2.9    &    0.90    &   0.003    & 3.0 & 3.0 & 3.0 \\
 \botrule
 \end{tabular}
 \end{table}

\begin{figure}[p]
 \includegraphics[width=0.8\textwidth]{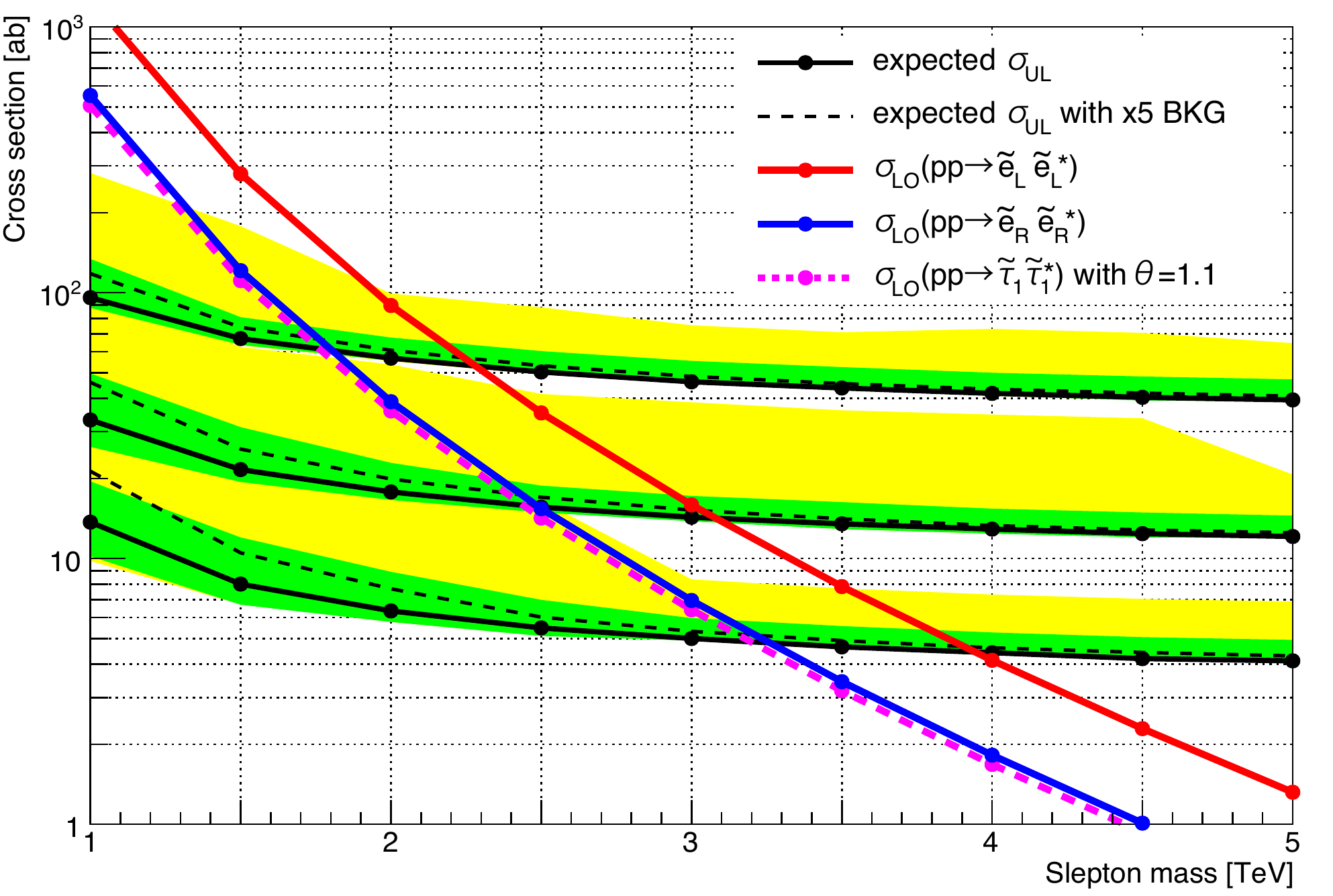}
 \caption{As in \figref{Limit14}, but for a $100\TeV$ $pp$ collider.
   The upper limits on the signal cross section (black solid lines
   with statistical uncertainty bands) are for integrated luminosities
   of $\int\mathcal L=0.3$, 1, and $3\iab$ from top to bottom.}
 \label{fig:Limit100}
\end{figure}

In~\figref{Limit100}, we show the resulting limits for different
scenarios.  The efficiency factors $\epsilon_\mu$ and
$\epsilon_{\tilde l}$, are taken into account in this plot.  The upper
bound $\sigma\s{UL}$ on the signal cross section is computed for
$\int\mathcal L=0.3$, 1, and $3\iab$ and is shown as black solid
lines.  The effect of statistical and systematic uncertainties are
displayed by the bands and the black dashed lines, respectively
(see~\figref{Limit14}).  The red (blue) line gives the production
cross section for a left- (right-) handed slepton, and the magenta
line corresponds to a slepton with a mixing angle of $\theta=1.1$,
which minimizes the production cross section.

\begin{table}[t]
 \caption{Expected mass limits of long-lived sleptons at the $14\TeV$
   LHC and future $100\TeV$ $pp$ colliders, based on the analysis
   described in \secsref{14}{100}; i.e., long-lived sleptons below
   these bounds are expected to be excluded at $95\%$ CL if no excess
   is observed.  Left-handed (right-handed) sleptons correspond to
   left--right mixing angles of $\theta=0$ ($\pi/2$), and ``least
   production'' is for the minimal signal cross section at
   $\theta=1.1$.  For $N\s{gen;LL}=3$, all sleptons are assumed to
   have the same mass and mixing angle. }
 \label{table:explimit}
 \begin{tabular}[t]{l|ccc|ccc}\toprule
  &\multicolumn{3}{|c|}{$14\TeV$ LHC}&\multicolumn{3}{|c}{$100\TeV$ $pp$ collider}\\
   & 0.1$\iab$ & 0.3$\iab$ & $3\iab$ & 0.3$\iab$ & $1\iab$ & $3\iab$ \\\hline
  $N\s{gen;LL}=1$, left-handed
  & $0.66\TeV$ & $0.83\TeV$ & $1.21\TeV$ & $2.28\TeV$ & $3.08\TeV$ & $3.95\TeV$ \\
  $N\s{gen;LL}=1$, right-handed
  & $0.55\TeV$ & $0.70\TeV$ & $1.07\TeV$ & $1.81\TeV$ & $2.49\TeV$ & $3.25\TeV$ \\
  $N\s{gen;LL}=1$, least production
  & $0.54\TeV$ & $0.69\TeV$ & $1.06\TeV$ & $1.76\TeV$ & $2.44\TeV$ & $3.20\TeV$ \\
\hline
  $N\s{gen;LL}=3$, all left-handed
  & $0.83\TeV$ & $1.01\TeV$ & $1.41\TeV$ & $3.02\TeV$ & $3.97\TeV$ & $4.96\TeV$ \\
  $N\s{gen;LL}=3$, all right-handed
   & $0.70\TeV$ & $0.88\TeV$ & $1.27\TeV$ & $2.45\TeV$ & $3.30\TeV$ & $4.20\TeV$ \\
\botrule
 \end{tabular}
\end{table}

 \section{Discussion}\label{sec:disc}

The results of our analysis are collected in~\tableref{explimit},
where we show the expected sensitivity of the $14\TeV$ LHC and $100\TeV$
collider.  The various entries show the lower bounds on long-lived
sleptons, assuming that the obtained data is consistent with the SM
expectation.  The first three lines of the table are based on pair
production of a single slepton type and denoted by $N\s{gen;LL}=1$.
The last two lines, with $N\s{gen;LL}=3$, assume three degenerate
long-lived sleptons, which are either left-handed or right-handed, so
that the production cross sections are a factor of three larger.

\begin{figure}[p]
 \includegraphics[width=0.7\textwidth]{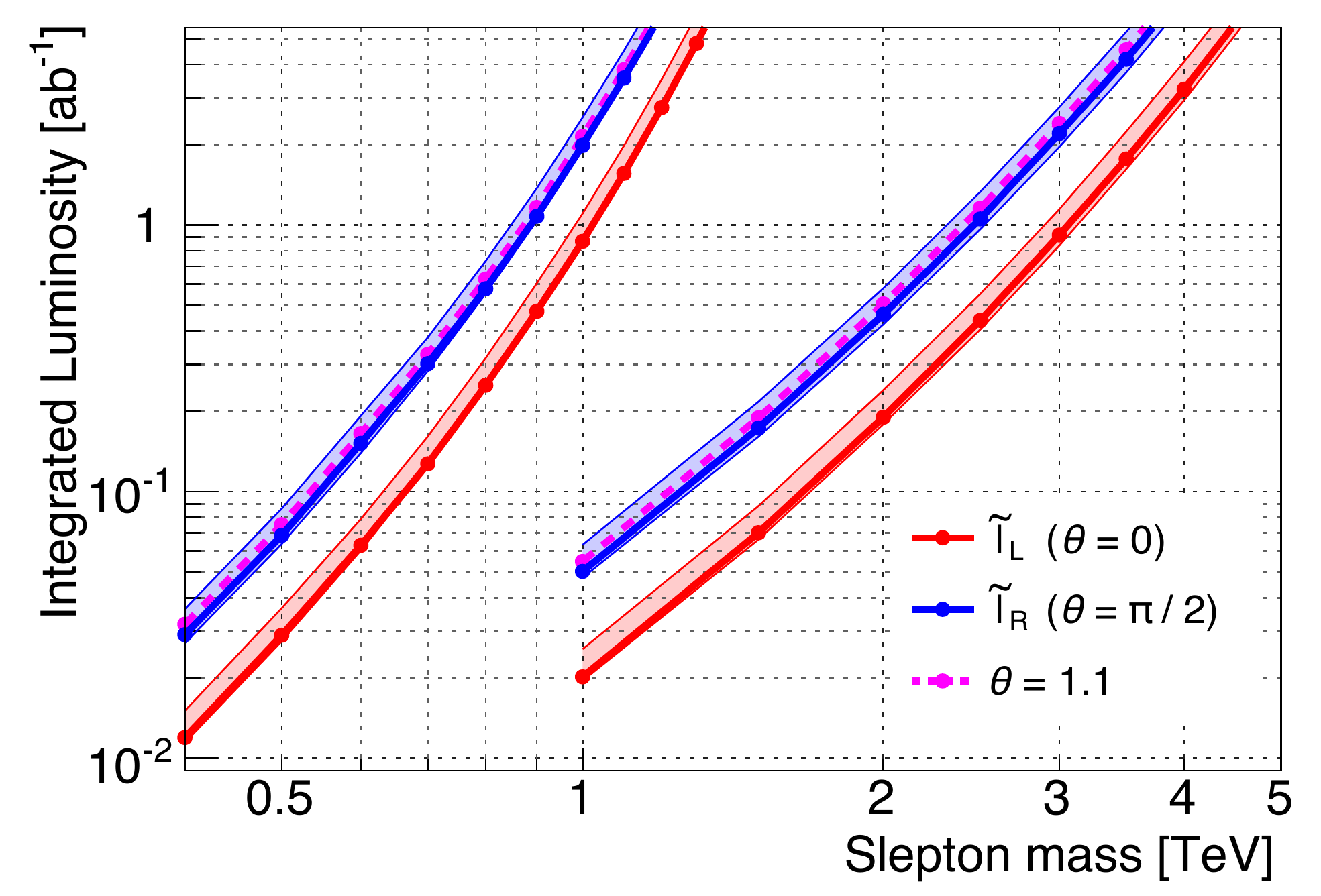}
  \caption{The integrated luminosity required for
    excluding long-lived sleptons at $95\%$ CL for $N\s{gen;LL}=1$ for
    three values of the slepton left--right mixing angle $\theta$. For
    each $\theta$, the left-hand (right-hand) contour shows the
    required integrated luminosity for the $14\TeV$ LHC (future
    $100\TeV$ $pp$ collider). The lines for $\theta=0$ and $\pi/2$ are
    drawn as bands, which show the $68\%$ statistical uncertainty of
    the expectation.}
 \label{fig:lumiplot}
\end{figure}

It is instructive to interpret these results in terms of the
integrated luminosity required for excluding long-lived sleptons of a
particular mass.  This is shown in \figref{lumiplot} for the case
$N\s{gen;LL}=1$ for $\theta=0$, $\pi/2$ and $1.1$.  For different
values of $N\s{gen;LL}$, the required luminosity is a factor of
$N\s{gen;LL}$ smaller.

\begin{figure}[p]
 \includegraphics[width=0.7\textwidth]{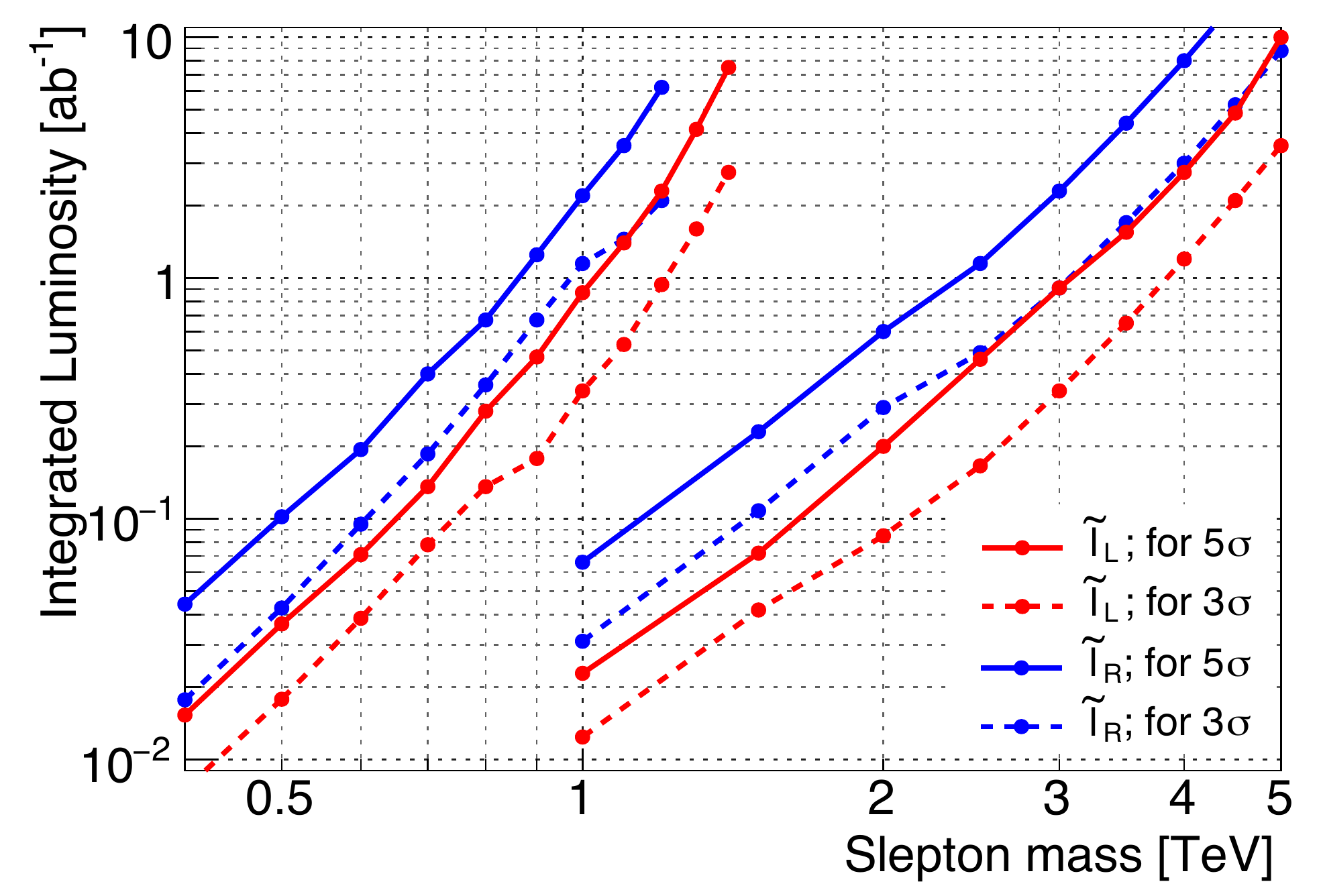}
 \caption{The same as \figref{lumiplot}, but for the discovery of
   long-lived sleptons. Solid (dashed) lines are for $5\sigma$-
   ($3\sigma$-)discoveries.}
 \label{fig:discoveryplot}
\end{figure}

In the same manner, discovery sensitivity, i.e., the luminosity
required for $3\sigma$-evidence and $5\sigma$-discovery, is
illustrated in \figref{discoveryplot}.  Here, the $5\sigma$-level
($3\sigma$-level) signature in one-sided tests is defined as having
the $p$-value of the background-only hypothesis smaller than
$2.9\EE-7$ $(0.0013)$.


 \begin{figure}[p]
\includegraphics[width=0.6\textwidth]{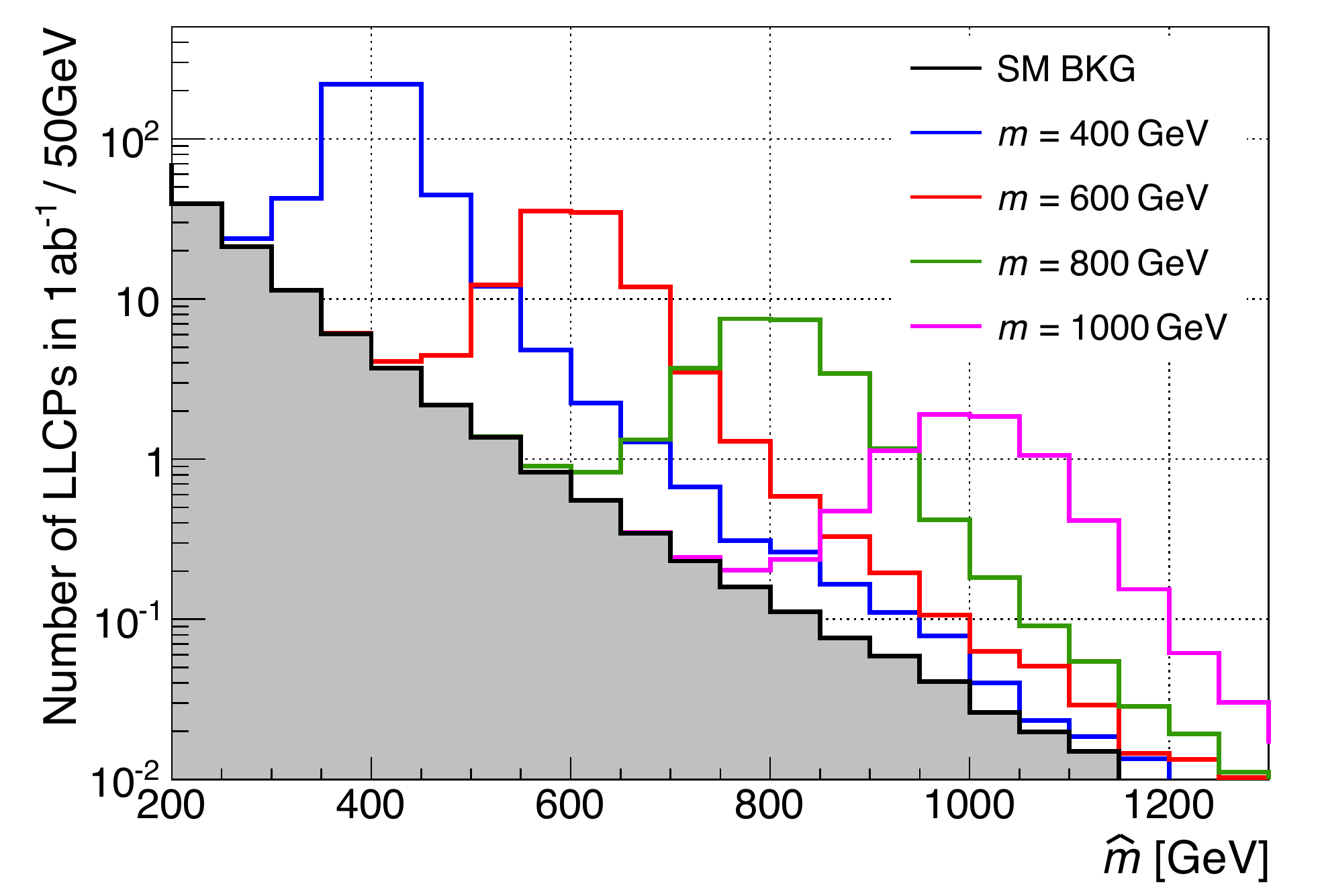}
  \caption{Distribution of the reconstructed LLCP mass $\hat m$ in the
    $14\TeV$ LHC analysis (\secref{14}). The gray region is the
    contribution from SM background muons, on which signal
    contributions are stacked.  Note that this shows the number of
    particles, not of events.  }
  \label{fig:mass14}
 \end{figure}

 \begin{figure}[p]
\includegraphics[width=0.6\textwidth]{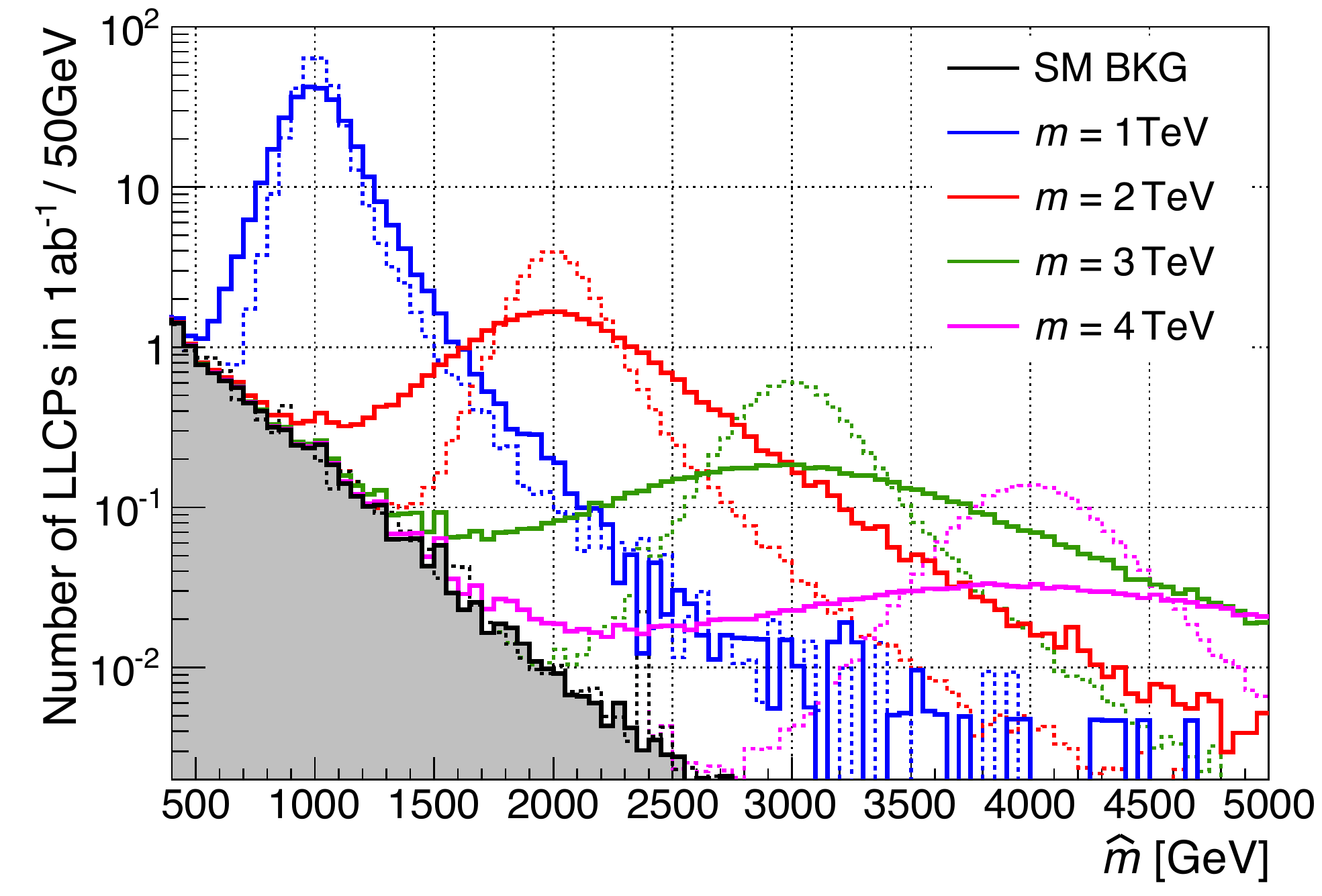}
  \caption{As in \figref{mass14}, but for the $100\TeV$ $pp$ collider
    analysis (\secref{100}). Solid lines are for the nominal analysis
    with $C=0.1\un{TeV^{-1}}$, and dotted lines show the result
    without momentum re-smearing, i.e., with $C=0$.}
  \label{fig:mass100}
\end{figure}
If an LLCP is discovered, the resolution of its mass determination
will be of great interest.  
\figsref{mass14}{mass100} display the reconstructed mass of LLCP candidates
in selected events.
As the slepton mass $m\s{LLCP}$ increases, the cross section decreases
and with it the number of true LLCPs produced.  At the same time,
because $\hat m$ peaks near $m\s{LLCP}$, the background contribution
under the peak falls sharply, and the $\hat{m}$ distribution in this
region is virtually background free.  It is for this reason that the
expected $\sigma\s{UL}$ in~\figsref{Limit14}{Limit100} is nearly flat
for larger $m\s{LLCP}$.

A key ingredient in the mass measurement is the resolution of the
momentum measurement, which typically deteriorates for large $\PT$.
In~\figref{mass100}, we examine the effect of the $\PT^2$ term
of~\eqref{momres} on the mass measurement.  In addition to the solid
lines obtained as described above with $C=0.1\un{TeV^{-1}}$, we show
the results obtained with $C=0$ as dotted lines.  Even with $C=0$, the
peaks are softer for larger values of the mass because the momentum
resolution scales as $\Delta\PT=0.05\PT$.  With non-zero $C$, the
resolution clearly deteriorates for $m\s{LLCP}=3\text{--}4\TeV$.
Thus, momentum resolution is crucial for the discovery of LLCPs with
masses $\agt 3\TeV$. It is also notable that the background
distribution (dotted black line) is hardly affected by this factor,
since it essentially cuts off below $3\TeV$.

Pile-up events are not included in this analysis, because the LLCP
searches focus on particles with very large momenta.  Pile-up events
tend to produce less energetic particles, so their effects on the
$\PT$ or $\beta$ measurements of very energetic particles should be
small.  On the other hand, pile-up may worsen the resolution of the
$E\s{loss}$ measurement, which we used to reduce background.  This
issue is related to lepton identification, and it should be carefully
examined in future studies on detector design.

For the $100\TeV$ analysis, we required $E\s{loss}$ below $30\GeV$,
since the typical energy loss of $\beta=0.4$ sleptons is around
$13\GeV$.  The $E\s{loss}$ cut reduces 18\% of fake LLCPs
(cf.~\tableref{selectionflow-100}), which ultimately reduces the
background events by 34\%, because signal events are required to have
two LLCPs.  If the energy resolution in the calorimeters is better
than assumed here, a tighter cut on $E\s{loss}$ could be used. On the
other hand, pile-up events may worsen the energy resolution.

\section{Conclusions}\label{sec:concl}

We have discussed the prospects for LLCP searches at the $14\TeV$ LHC
and at a $100\TeV$ $pp$ collider.  We use sleptons as the benchmark
LLCP, with Drell--Yan slepton-pair production as the sole slepton
production channel.

For scenarios in which only a single type of long-lived slepton is
produced ($N\s{gen;LL}=1$), the $14\TeV$ LHC is expected to constrain
the LLCP mass as $m\s{LLCP}>700\text{--}800\GeV$ with $0.3\iab$, and
$1.1\text{--}1.2\TeV$ with $3\iab$, depending on the amount of
left--right slepton mixing.  Thus, the entire parameter space of the
slepton--neutralino co-annihilation scenario can be probed at the LHC.
At a $100\TeV$ $pp$ collider, the sensitivity is expected to reach
1.8--$2.3\TeV$ for $0.3\iab$, and 3.2--$4.0\TeV$ for $3\iab$.  In
terms of discovery, the $14\TeV$ LHC is expected to discover
$600\text{--}800\GeV$ ($1.0\text{--}1.2\TeV$) long-lived sleptons with
$0.3\iab$ ($3\iab$), while a $100\TeV$ collider's coverage is
estimated to slepton masses up to $1.6\text{--}2.2\TeV$
($3.1\text{--}4.1\TeV$) with $0.3\iab$ ($3\iab$).

We have found that, in $100\TeV$ proton collisions, the radiative energy loss of
energetic muons is significant.  We exploited
this fact to reject fake LLCPs coming from SM muons.  On the other
hand, the momentum resolution, which plays a key role in the LLCP mass
measurement, will be more challenging at a high-energy collider.  This
effect is clearly seen in~\figref{mass100}.  The momentum resolution
can be improved by increasing the magnetic field strength as well as
by using a bigger tracker.  Since momentum measurements are essential
for any searches at collider experiments, detailed studies of the
required resolution and the implications for detector design are
critical.

\section*{Acknowledgments}

J.L.F.\ and Y.S.\ thank the CERN Theoretical Physics Group and
J.L.F.\ thanks the Technion Center for Particle Physics for
hospitality.  The work of J.L.F.\ and Y.S.\ is supported in part by
BSF Grant No.~2010221.  The work of J.L.F.\ is also supported in part
by NSF Grant No.~PHY--1316792 and by a Guggenheim Foundation grant.
The research of Y.S., S.I., and S.T.\ is also supported by the ICORE
Program of Planning and Budgeting Committee, and by ISF Grant
Nos.~1937/12 (Y.S.\ and S.I.) and 1787/11 (S.T.).

\appendix

\section{Detailed Description of the Monte Carlo Simulation}

\subsection{Background Events}

We use the Snowmass backgrounds for $pp$ colliders with $\sqrt s=14$
and
$100\TeV$~\cite{Anderson:2013kxz,Avetisyan:2013onh,Avetisyan:2013dta}
as the SM background contributions.  These backgrounds are available
with and without pile-up; for simplicity, we used the backgrounds
without pile-up.  Here we review the procedure used to generate the
Snowmass background.

The backgrounds were generated with
\software{MadGraph\,5}~\cite{MADGRAPH5}, together with
\software{BRIDGE}~\cite{BRIDGE}.
\software{Pythia\,6.4}~\cite{Pythia6.4} was used for parton showering
and hadronization with the \software{Pythia-PGS} interface, and
\software{Delphes}, tuned by the Snowmass Collaboration based on
\software{Delphes\,3.0.9}~\cite{deFavereau:2013fsa,Cacciari:2011ma,Cacciari:2005hq},
was used for detector simulation, with jet reconstruction implemented
with \software{FastJet}~\cite{Cacciari:2005hq,Cacciari:2011ma}.

The detector simulation of the background events, which is summarized
in Ref.~\cite{Anderson:2013kxz}\footnote{The parameters are slightly
  modified: {\tt Radius} in {\tt ParticlePropagator} is set to
  $1.29\un{m}$, and the muon tracking efficiency is set to 99.98\% for
  $|\eta|<1.5$ and 98.0\% for $1.5<|\eta|<2.5$.}, is based on a
detector which has a tracker, an electromagnetic and a hadronic
calorimeter.  Tracking efficiency and resolution in the tracker, and
energy resolution in the calorimeters are included.  The energy flow
method is utilized for calorimeter analysis.

Electrons ($e^\pm$) and muons ($\mu^\pm$) are reconstructed from an
isolated track originating from true electrons and true muons,
respectively, where a charged-particle track is called isolated if the
scalar sum of $\PT$ of tracks and calorimeter hits around the track
($\Delta R<0.3$) is less than $10\%$ of the track $\PT$.  Electrons
(muons) must satisfy $\PT>10\GeV$ and $|\eta|<2.5$ (2.4).  Lepton
momentum is smeared with a tracker resolution, which for muons with
$\PT>200\GeV$ is set to $\Delta\PT=0.05\PT$.  Note that the
information from the calorimeters is not used here, and that
misidentifications are not considered.

Jets are reconstructed by the anti-$k_{\rm T}$ algorithm with $\Delta
R=0.5$ using the \software{FastJet} package.  A calorimeter cluster is
identified as a photon if the cluster has hits from photons or
$e^\pm$'s and $\mu^\pm$'s, but is not associated with any
reconstructed track.  Otherwise the cluster is identified as a jet.
The missing energy ($\mET$) is calculated from the reconstructed
tracks and calorimeter hits.

\subsection{Signal Events}

Signal events are generated following the procedure of Snowmass
background generation.  \software{Madgraph\,5}~\cite{MG5AMC} is used
as the event generator, and \software{Pythia\,6.428} and
\software{Delphes} are used with the same parameters used to generate
the Snowmass background.

The long-lived sleptons are treated as stable particles. They are
reconstructed as muons ($\mu^\pm$) if they have velocity $\beta>0.3$,
$\PT>10\GeV$ and $|\eta|<2.4$; otherwise they are ignored.

\subsection{Momentum Re-smearing}

As discussed in \secref{100}, momentum resolution deteriorates at very
high momentum, because the trajectory becomes straighter for large
$\PT$.  For very large $\PT$, the momentum resolution is approximated
as $\Delta\PT\propto\PT^2$.  This effect is important for a $100\TeV$
collider, but it has not yet been modeled in the procedure described
above.  Therefore, for the $100\TeV$ collider simulation we smeared
the reconstructed momentum of charged tracks again with the
distribution $\mathop{\mathrm N}(\PT, C\PT^2)$, with
$C=0.1\un{TeV^{-1}}$.  In the $14\TeV$ simulation, this re-smearing
was not employed.

\subsection{Object Selection}

The background events provided by the Snowmass collaboration and the
generated signal events are then subjected to further object
selections.  First, all objects with $\PT<100\GeV$ in the $100\TeV$
analysis, and with $\PT<30\GeV$ in the $14\TeV$ analysis, are removed.
Electrons, jets, and photons are required to have $|\eta|<2.5$, and
muons are required to have $|\eta|<2.4$.  Muon pairs are removed if
their invariant masses satisfy $|m_{\mu\mu}-m_Z|<5\GeV$.

Then, a ``muon'' is regarded as a LLCP if it satisfies the following
conditions:
\begin{itemize}
 \item $\Delta R>0.5$ from the nearest objects (electrons, muons,
   jets, and photons)
 \item In the $14\TeV$ analysis, $\hat\PT>100\GeV$, $|\eta|<2.4$, and
   $0.3<\hat\beta<0.95$
 \item In the $100\TeV$ analysis, $\hat\PT>500\GeV$, $|\eta|<2.4$, and
   $0.4<\hat\beta<0.95$
 \item In the $100\TeV$ analysis, $\Eloss < 30\GeV$\, .
\end{itemize}
In the conditions above, $\hat\beta$ is the smeared velocity, which
obeys the following distributions
\begin{align}
 {\mathop{\mathrm{PDF}}}(\hat\beta)_\mu
  &= 0.832\mathop{\mathrm N}(1,0.022)
  + 0.162\mathop{\mathrm N}(1,0.050)
  + 0.00534\mathop{\mathrm N}(1,0.116),\\
{\mathop{\mathrm{PDF}}}(\hat\beta^{-1})_{\tilde l}
  &= \mathop{\mathrm N}(\beta^{-1},0.025) \, ,
\end{align}
for background muons and signal sleptons, respectively. $\Eloss$ is
the energy deposit of the particles in the calorimeter.  For muons, it
is simulated with \software{Geant\,4.10}~\cite{GEANT4}, where the
calorimeter is approximated as $3.0\un{m}$ iron.  Sleptons are assumed
to pass this cut because the energy loss is far less than the
threshold (see \secref{100}).

After these object identifications, events are selected and analyzed
as summarized in \secsref{14}{100}.

\bibliography{bibsleptons100TeV}

\end{document}